\documentclass[aps,twocolumn, pre,superscriptaddress, longbibliography, notitlepage]{revtex4-2}
\usepackage{bm,amsmath,amssymb,graphicx, mathtools,multirow,gensymb}
\usepackage{svg}
\usepackage[colorlinks=true,linkcolor=MidnightBlue,urlcolor=black,citecolor=MidnightBlue,anchorcolor=MidnightBlue]{hyperref}\usepackage[dvipsnames]{xcolor}

\begin{document}
\title{Hydrodynamic Phase Separation and Morphological Evolution in Chiral Active-Passive Mixtures}
\author{Mayurakshi Deb}
\email{ph22d091@smail.iitm.ac.in}
\affiliation{Department of Physics, Indian Institute of Technology Madras, Chennai, India}
\author{Rajesh Singh}
\email{rsingh@physics.iitm.ac.in}
\affiliation{Department of Physics, Indian Institute of Technology Madras, Chennai, India}
\begin{abstract}
The collective behavior of passive particles within chiral active matter has emerged as a significant area of soft matter research. However, most existing studies focus on systems where chirality is imposed by external torques rather than intrinsic activity. In this work, we 
study emergent dynamics
in a 
suspension of 
active spinners and passive colloids
by computing many-body hydrodynamic interactions via Ewald summation. By systematically exploring a broad range of area fractions and rotational velocities, we identify distinct phase-separation regimes sensitive to the system's kinematic parameters. Specifically, we report the emergence of unique structural morphologies, including the formation of passive particle vortices surrounding phase-separated active spinners and the development of large-scale active-passive bands. We characterize the underlying dynamics by analyzing the temporal evolution of characteristic length scales and the non-equilibrium velocity distributions of the passive particles. 
Our findings provide new insights into the role of long-range hydrodynamic couplings in governing the self-organization of non-equilibrium condensed matter.
\end{abstract}
\maketitle
\section{Introduction}
Active matter systems, composed of individual units that consume internal or ambient energy to generate self-propulsion, have revolutionized our understanding of non-equilibrium statistical mechanics \cite{vrugt2025exactly, marchetti2013, bowick2022}. From biological swarms and bacterial colonies to synthetic Janus particles and micro-rollers, these systems exhibit emergent collective behaviors—such as motility-induced phase separation   and giant number fluctuations—that are forbidden in equilibrium systems \cite{cates2015, cates2025active, ramaswamy2010, di2010bacterial}. 
While much of the foundational work in active matter has focused on achiral self-propelled particles (SPPs) that move along linear trajectories, recent attention has shifted toward ``chiral active matter," where the constituents exhibit a permanent preference for clockwise or counter-clockwise rotation \cite{lowen2016chirality}.
Chirality at the microscale is ubiquitous in biological systems, seen in the helical motion of sperm cells, the beating of cilia, and the circular trajectories of certain microorganisms near surfaces \cite{lauga2020fluid}.  
In these systems, the interplay between rotational activity, hydrodynamic interactions, and steric interactions leads to novel phenomena, including edge currents, odd viscosity, and the formation of living crystals \cite{tan2022odd, petroff2015fast, han2021fluctuating, singh2016crystallization}.
\\

Exploring collective behavior in active-passive mixtures represents one of the most compelling frontiers in modern active matter research
\cite{angelaniPRL2011, stenhammar2015activity,krafnickPRE2015,
yeo2016dynamics, smrek2017small, 
dolai2018phase,reichhardt2019reversibility,
liuPRL2020, mousavi2021active, 
madden2022hydrodynamically, gokhale2022dynamic, 
grober2023unconventional,
kushwaha2023phase, 
yuan2024colloid, garcia2025dynamics,serna2025,
jhajhria2025kinetics, bhattacharyya2025active, chamolly2026self}.
In such active-passive mixtures, the 
active component acts as a non-equilibrium bath, 
imparting ``active stress" onto the passive particles in the system leading to large-scale self-organization. 
Previous studies of chiral active systems using dry models (neglecting fluid-mediated interactions) have shown that active spinners can induce effective attractions between passive particles \cite{nguyen2014emergent}. However, the role of the solvent remains a topic of sustained research. In the low-Reynolds-number regime characteristic of microfluidic environments, particles are coupled through long-range, many-body hydrodynamic interactions (HI) \cite{lauga2020fluid}. 
These interactions are not merely perturbations; they fundamentally alter the stress distribution within the fluid and can either stabilize or disrupt the structural motifs predicted by dry models. 
All the studies incorporating HI, this far, 
have been dedicated to study phase separation of colloids which are rotating due to to a net-torque on them \cite{yeo2015collective,massana2021arrested,hrishikesh2022phase,yuan2024colloid, yeo2016dynamics, lushi2015periodic}. 
Crucially, flow due to particles driven by a net torque decays as $1/r^2$ (where $r$ is the distance from the source), while the flow due to an active spinner 
decays as $1/r^4$ \cite{ghose2014irreducible, ishikawa2020stability}. 
To the best of our knowledge, a study of collective dynamics in a system of chiral active particles, where an isolated particle can actively spin is lacking (an investigation on two-body dynamics of rotors: one active and one passive rotor is present in \cite{fily2012cooperative}).

In this work, we address this gap by
studying a mixture of truly active spinners and passive colloids.
We demonstrate that the injection of rotational energy at the microscopic level triggers a dramatic breakdown of the uniform state. 
We explore the morphological evolution of the phase separation across a wide parameter space. Hydrodynamic interactions are fully accounted for by computing the Stokesian flow fields produced by the active spinners, as we describe below. 
Our investigation reveals that the fluid-mediated coupling between 
active spinners and passive 
colloids drives the system into striking morphologies, ranging from localized ``vortices" of passive particles around phase-separated active particles to macroscopic bands of active and passive particles. The structures are 
determined primarily by
the intensity of the chirality of the active particles and
the density of the mixture.
Our findings highlight a distinct mechanism for phase separation and demonstrate how chiral activity plus long‑range fluid flow can produce organized, large‑scale dissipative structures with possible relevance for designing new active materials and understanding collective behavior in biological systems.

\begin{figure*}[t]
    \centering
    \includegraphics[width=0.99\textwidth]{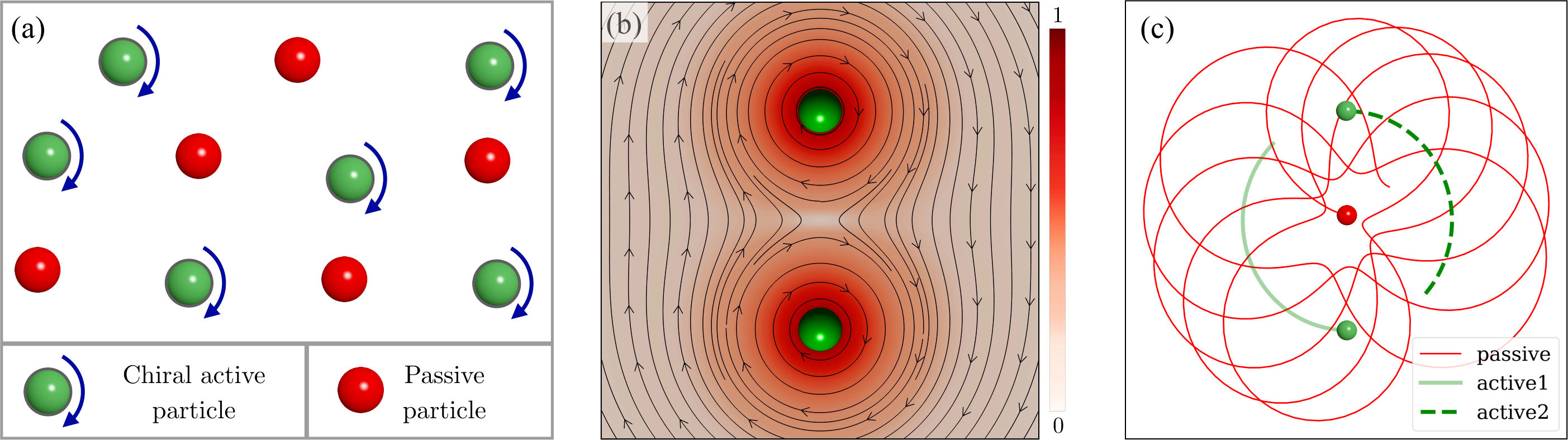} 
    \caption{
    (a)
 Schematic representation of the system: spinning active particles  are shown in green with curved arrows, while red circles are passive particles. Panel (b) contains the streamlines of
the fluid flow due to two active spinners. The
flow streamlines have been overlaid on the normalized flow speed. 
Panel (c) shows trajectories of two active (green) and one passive (red). Green particles revolve around each other, while the passive particle (shown in red) periodically moves around active particles when placed slightly away from the center of line joining the active particles. 
    }
    \label{fig:schematic}
\end{figure*}
 
The rest of the paper is organised as follows. In Section \ref{sec:model}, we present our theoretical model and methodology. We also provide details of dimensionless parameters of the system. In section \ref{sec:phase} we discuss the behaviour of these suspensions at different values of the spinning speed and the overall area fraction.  In section \ref{sec:coarsenig} we discuss the coarsening dynamics of the system and section \ref{sec:vel_dist} discusses the probability distribution of velocities of the passive particles.In section \ref{sec:dyn_passive}, we infer the phenomenon at the interface of actives and passives from displacement plots. In section \ref{subsec:asymm} we discuss the role of number asymmetry in the system.
Finally, we conclude the work in \ref{sec:summary} 
by summarising our results and suggesting future directions.

\section{Model}\label{sec:model}
\subsection{An active particle as a sphere with a slip}
{In this section, we describe our model for active particles.
We model an active particle
as a spherical particle with a surface slip $\boldsymbol{v}^{\mathcal{A}}$ 
\cite{anderson1989colloid,ebbens2010pursuit}. 
Fluid velocity on the surface of active particle is then:
\begin{equation}
    \boldsymbol{v}(\mathbf{x}_\alpha+\boldsymbol{b}_\alpha)=\mathbf{U}_\alpha+\boldsymbol{\Omega}_\alpha\times\boldsymbol{b}_\alpha+\boldsymbol{v}^{\mathcal A}_\alpha(\boldsymbol{b}_\alpha).
    \label{eq:slip}
\end{equation}
The slip velocity $\boldsymbol{v}_\alpha^{\mathcal A}$ originates from non-equilibrium processes at the colloidal interfaces. This may arise from ciliary motion in microorganisms \cite{lighthill1952,blake1971a,brennen1977} or from gradients in chemical potential (or any other phoretic field \cite{anderson1989colloid, ebbens2010pursuit}) on the surface of a colloidal particle.
The active slip can take any arbitrary functional form, subject only to the constraint imposed by fluid mass conservation, i.e., $\int \boldsymbol{v}^{\mathcal A}_\alpha(\boldsymbol{b}_\alpha)\cdot \hat{\boldsymbol b}_\alpha\,\mathrm{d}\mathcal S_\alpha=0$. Here $\hat{\boldsymbol b}_\alpha = \boldsymbol{b}/b$ is the unit vector along the radius of the $\alpha$-th particle, while $b=|\boldsymbol{b}|$. In this paper, we use Greek alphabets, such as $\alpha,\beta$ to denote particle indices, while Cartesian indices are written using Roman alphabets $i,j$, etc.
To compute hydrodynamic interactions between active particle with surface slip,
we use known results from a theory which is based on expanding the slip in an irreducible basis 
\cite{singh2015many,singh2018generalized,turk2022stokes, singh2016crystallization,turk2024fluctuating, turk2025autophoretic, deb2025ewald}. The method gives contributions due to each irreducible mode of the slip.  In this paper, we only retain the $4a$ mode of the slip, which corresponds to self-spinning of an active particle. 
The $4a$ mode corresponds to 
a chiral octupole of active slip \cite{singh2015many, turk2022stokes, nemeth2026nonreciprocal}. 
Such a mode creates a flow which decays as $1/r^4$. The $4a$ mode leads to active spinning of a particle in absence of any net external torque on it \cite{ghose2014irreducible, pak2014generalized}. 
}


\subsection{Equations of motion for many active particles}
 A schematic diagram of our system is present in Fig.\ref{fig:schematic}. 
All the particles are initialised on a two-dimensional surface ($xy-$plane) 
and have their orientation along the $\hat z$ axis. We consider the orientation of the colloids to be fixed $\hat{\bm{e}} = \hat{\bm z}$. This is the limit of large bottom-heaviness of the colloid. Our model is motivated from the  
green algae \emph{Volvox} \cite{drescher2009,goldstein2015green} which is known to actively spin around its axis. The 
\emph{Volvox} is also known to be bottom-heavy which stabilizes its axis of rotation \cite{goldstein2015green}. 
The system has $N_p$ passives and $N_a$  active spinners. In sections \ref{sec:model} to \ref{sec:dyn_passive}, $N_a=N_p$.
The position $\bm x_{\alpha}$ 
of the $\alpha^\text{th}$ colloidal particle follows the following dynamical equation:
\begin{alignat}{1}
\dot{ \bm x}_{\alpha}  = {\mu}^T  \,\bm{F}_{\alpha}^{e} 
&+{\sum_{\alpha=1}^{N}}\sum'_n
  \Big(
  \bm U^{1s}_{\alpha\beta} +
  \bm U^{4a}_{\alpha\beta} 
  \Big).
\label{eq:mainEq}
\end{alignat}
Here, the sum $n$ represents lattice summations to implement periodic boundary conditions in a box of size $L$. 
The prime indicates that $\alpha=\beta$ is excluded in the main box $n=1$. It should be noted that the active contribution is only due to the $N_a$ active particles. Here,
$\mu_t =   \frac{1}{6\pi\eta b}-\frac{\xi}{\pi^{3/2}\eta }+\frac{20\xi^3b^2}{9\eta\pi^{3/2}},$ 
 where the contributions proportional to $\xi$ are due to self-contributions which stem from the Ewald summation procedure  \cite{beenakker1986ewald}, while they vanish in the limit of $L\rightarrow \infty$.
 The hydrodynamic interactions between particles - as given in Eq.\eqref{eq:mainEq} -  can be written explicitly as:
\begin{subequations}
\begin{align}
 \bm U^{1s}_{\alpha\beta} &= \left(  1 + \tfrac{b^2}{3} \nabla^{2} \right)\bm{G} (\bm x_\alpha, \bm x_\beta)\cdot \bm F^e_\beta\\
 \bm U^{4a}_{\alpha\beta} &=  
   \bm{\nabla}_{\beta} 
   \bm{\nabla}_{\beta} 
    \left[\bm{\nabla}_{\beta}\times\bm{G} (\bm x_\alpha, \bm x_\beta)
    \right]\cdot \bm \Gamma_\beta
\end{align}
\label{eq:U1s4a}
\end{subequations}
For explicit forms of $\bm U^{1s}_{\alpha\beta}$ and $\bm U^{4a}_{\alpha\beta}$ in index notation, see appendix \ref{seappc:explicitRBM}.  Here $\bm{G}$ is a Green's function 
In addition $\bm 
\Gamma_\beta$ is a symmetric-irreducible tensor of rank 3, 
which represent the chiral activity due to the $4a$ mode.
Note that this term is only non-zero for active particles. 
We parameterize it in terms of the orientation $\hat e$ of the swimmers: 
${\Gamma}_{ijk} =b^5\eta\omega_s \,\left(5\hat{e }_{i}\hat{e }_{j}\hat{e }_{k}-\left[\hat{e }_{i}\delta_{jk}+\hat{e }_{j}\delta_{ik}+\hat{e }_{k}\delta_{ij}\right]\right).$
Here, $\omega_s$ is a constant, which controls active swimming speed of the particles.
It is worthwhile to note that dynamics in our systems is solely due to the flow generated by the active spinners. There are no net external forces and torques in our 
system (stochastic terms are assumed to be subdominant to 
deterministic contributions, and thus, ignored). 
Although, body force $\bm{F}_{\alpha}^{e}$ is included to avoid overlap of particles. 
The body force on the $\alpha$th particle is given as: 
    $\bm F^e_\alpha = -\bm\nabla_\alpha\,  \mathcal U$, 
    where $\mathcal U=\sum_{\alpha<\beta}\mathcal U^e   (\bm x_\alpha,\bm x_\beta)$.
Here, $\mathcal U^e$ is a pairwise repulsive potential precluding overlap of particles from steric interactions. We choose it to be of the form: $\mathcal U^e=\tfrac12\,k^e\left(r_{\alpha\beta}-2b\right)^2$, if $r_{\alpha\beta}<2b$ and it vanishes otherwise.
Here, $k^e$ is a positive constant that determines the strength of the repulsive force due to steric interactions between the particles.
In the above, we have defined $r_{\alpha\beta}=|\bm x_\alpha - \bm x_\beta|$ and $b$ is the radius of the particles. 
Further details about the simulation and parameters used are given in appendix \ref{app:sim_details}. Next, we describe the key dimensionless parameters.

\subsection{Dimensionless Parameters}
In this section, we define the dimensionless numbers 
of the system and explain their physical significance. 
The overall area fraction $\phi$ is a number which  
is given by the ratio of area occupied by the particles (both active and passive) to the area of the box. 
The dimensionless active-spinning speed $\psi$  is the ratio of the spinning speed to the quantity $\omega_0$, which we explain below. Explicit forms of the parameters $\psi$ and $\phi$ are:
\begin{align}
  \psi = \frac{\omega_s}{\omega_0},\quad \omega_0 = \frac{k^e}{\eta b},\quad    \phi= N\pi\frac{b^2 }{L^2}
  \label{eq:dimParams}
     \end{align}
     Here, $N$ is the total number of particles and $b$ is the radius of the active and passive particles, while $k^e$ is the strength of repulsive potential, $\eta$ is the viscosity and $\omega_s$ is the spinning speed of active particle.
To elucidate the physical meaning of the dimensionless parameter $\psi$, it is useful to study the time scale $\tau$, which depends on the strength $k_e$ of the steric repulsion force and the viscous drag: 
$\tau = \frac{1}{\omega_0}= \frac{\eta b}{k^e}$. 
Thus, it controls the time scale at which the steric interaction, which precludes overlap of particles, becomes important. The competition between fluid flow driven by spinning of active particles and steric interactions between the particles plays a crucial role in the phenomenology we report. 
Thus, the ratio $\psi$ is a key dimensionless parameter which has been used to enumerate the phase diagram. 
\begin{figure}[b]
    \includegraphics[width=0.45\textwidth]{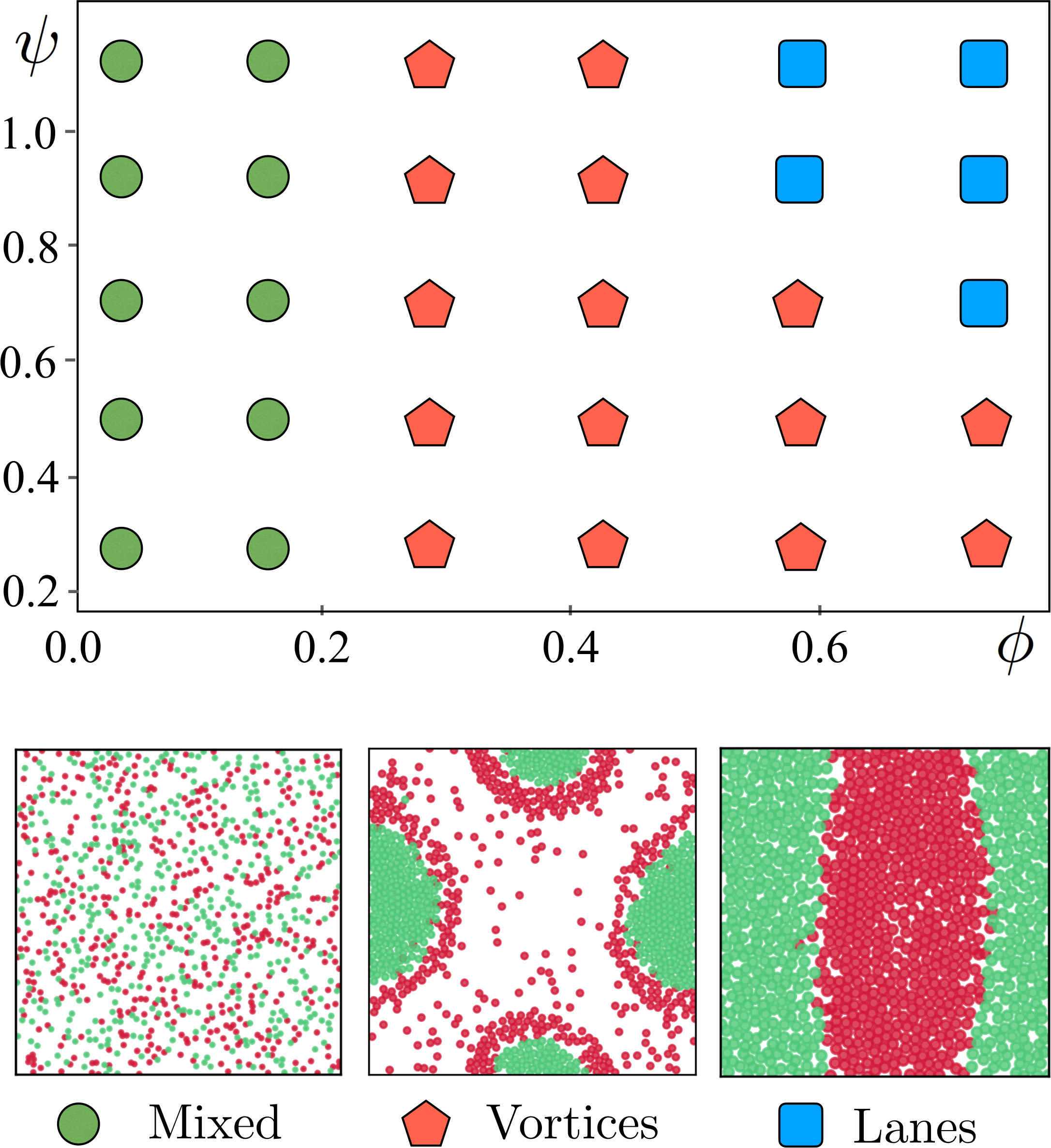}
    \caption{
    Top panel shows phase diagram in the plane of dimensionless
    active spinning speed
    $\psi$ and area fraction  $\phi$. See Eq.\eqref{eq:dimParams} for definition of these dimensionless parameters.
    Representative snapshots from each phase in the steady-state is shown along with key in the bottom panel. In the snapshots, 
    chiral active particles are shown in green,
    while passive particles are shown in red. See text for details.
\label{fig:phaseSep_4a}
}
\end{figure}

\section{Results}\label{sec:Res}
\subsection{Phase separation in a binary mixture of active and passive particles}
\label{sec:phase}
In this section, we present results from the study of phase separation in a suspension of passive and active particles using
the model and method of previous section. 
The results of simulation of the above model is shown in Fig.\ref{fig:phaseSep_4a}. We show that a suspension of active spinners and passive particles, phase separate into domains beyond a given spinning speed and area fraction. 
For area fractions up to $\sim 0.18$ the system remains in a homogeneous state and no phase separation is observed.
This phenomena can be understood from the fact that the particles are not close enough to interact with each other at these low area fractions. For area fractions of 0.2 and above, we observe that the system separates into active and passive domains. Active clusters surrounded by vortices formed by passive particles are observed at intermediate area fractions while bands are observed at very high area fractions and spinning speeds as seen in Fig. \ref{fig:phaseSep_4a}.
Crucially, we find the vortices phase can remain arrested in clusters at large values of activity. We described this in detail below. 
It is to be noted that, a suspension of active spinners alone  displays no clustering over time but the addition of the passive particles to  the suspension, gives rise to an effective attraction between the actives which now forms clusters. 
The 
role of number asymmetry between active and passive particles 
is studied further in section \ref{subsec:asymm}. See supplementary movie, as described in appendix \ref{supplemntary}, for the dynamical simulation of the system.

\subsection{Coarsening kinetics}
\label{sec:coarsenig}

To quantify the growth of domains in a 50:50 mixture of active and passive particles,
we study the length scale of the system during the coarsening phenomena. 
We compute the characteristic length scale 
from the average of the magnitude $k$ of the wave-vector in our system. 
Explicitly, the length scale $L(t)$
is computed using the expression 
\cite{kendon2001inertial}:
%

\begin{figure}[t]
\includegraphics[width=0.46\textwidth]{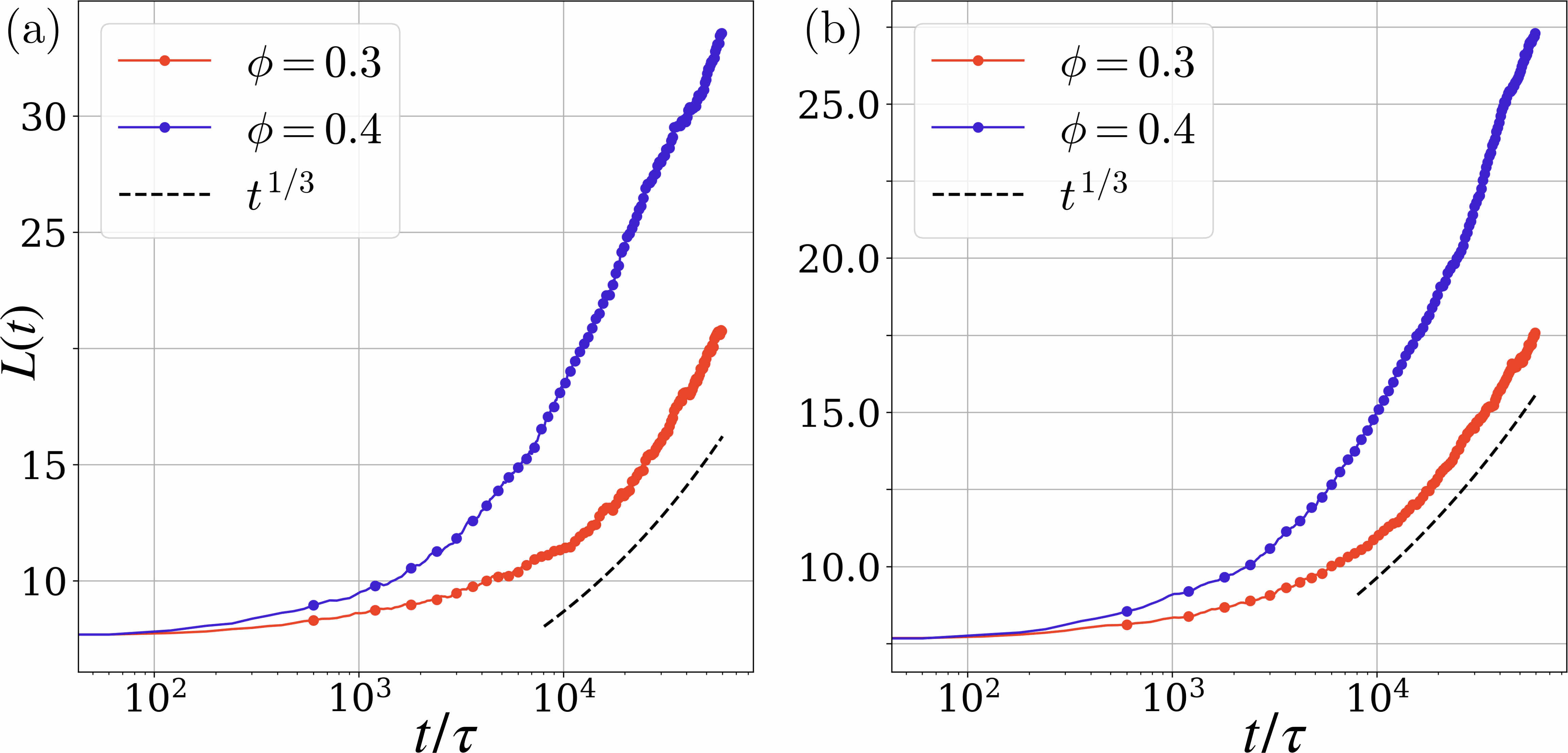}
 \caption{Coarsening kinetics during phase separation in a system of 5000 particles, where $N_a=N_p=2500$, for two different values of area fraction $\phi=0.3$ and $\phi=0.4$ at $\psi=1.2$ (in panel a) and $\psi=0.4$ (in panel b). 
 We quantify the growth of 
  length scale $L(t)$ of the coarsening domain using Eq.\eqref{eq:avgK}.  
 \textcolor{black}{Dotted line shows a growth exponent of $1/3$ for coarsening domain for comparison.} 
 }
    \label{fig:Lvst}
\end{figure}
%
\begin{align}
L(t)=\frac1b \frac{2\pi}{\left\langle k(t)\right\rangle}
,\qquad
 \left\langle k(t)\right\rangle
 = 
 \dfrac{\int 
 k S(k,t)\, dk}{\int S(k,t)\, dk}.
 \label{eq:avgK}
\end{align}
Here $S(k,t)$ is the spherically averaged structure factor. 
We compute structure factor by creating a density field from the location of the particles. In the above, we divide by 
the radius $b$  of the particles such that we 
measure a dimensionless length scale $L(t)$. 
As observed from the plot of $L(t)$ as a function of time $t/\tau$ in Fig.\ref{fig:Lvst}, 
\textcolor{black}{
the length scale of the coarsening domain appears 
to grow
with a growth exponent of $1/3$ at early-times. 
Although, there
are some deviations
in the simulation with respect to the 1/3 growth law.}
\textcolor{black}{A detailed analysis to find the asymptotic growth law
using finite-size analysis suggests an exciting direction
for future work.}
Indeed, the exponent $1/3$ corresponds to the model B dynamics \cite{hohenberg1977theory, bray1994theory}, which is 
governed solely by diffusion-limited kinetics.
At the early times, the 
phase separation is 
occurring due to the exchange of particles 
at the interface of actives and passives, while the total number of active and passive particles in the system is conserved. 
\textcolor{black}{
Thus, at early-times, a diffusive growth is possible.}
We note that an exponent of 1/3 for the growth of length scale during coarsening has been earlier reported for suspensions of colloidal particles with hydrodynamic interactions \cite{SIEROU_BRADY_2004, yeo2015collective}.

\subsection{Velocity distributions of passive particles}
\label{sec:vel_dist}
\begin{figure}[b]
    \centering
    \includegraphics[width=.45\textwidth]{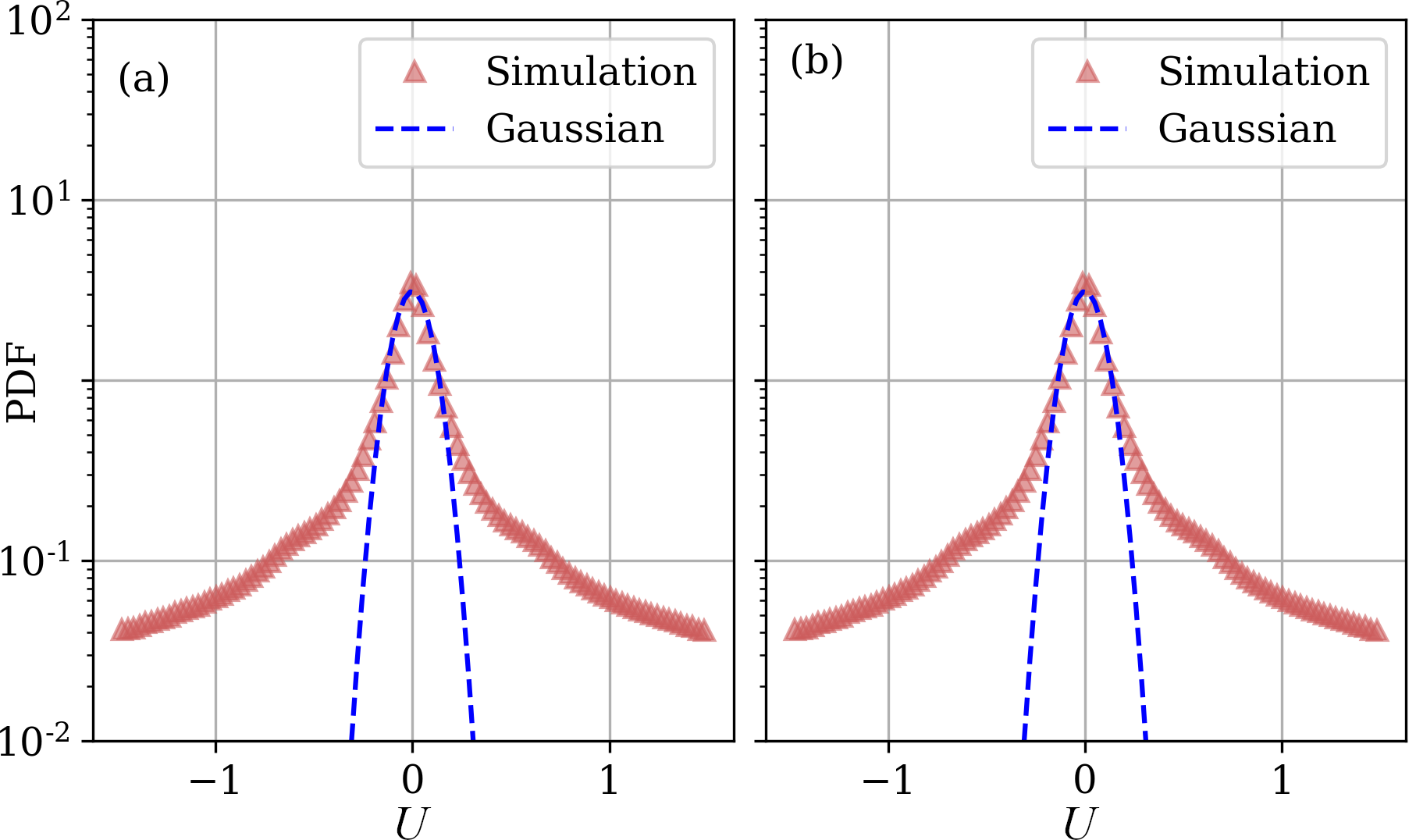}
    \caption{PDF (probability distribution function) for the $x$-component of the velocity ($U$) for passive particle. 
    A similar graph is obtained for the $y$-component. A best fit to the Gaussian distribution is shown using dotted lines. 
    We find that the PDF is strongly non-Gaussian. In the figure above, the panels are given as (a) $\phi=0.3$, $\psi=0.4$, (b) $\phi=0.4$, $\psi=0.4$. 
    Total number of particles is 5000, with $N_a=N_p=2500$}
    \label{vel_dist}
\end{figure}

\begin{figure*}[t]
    \includegraphics[width=0.999\textwidth]{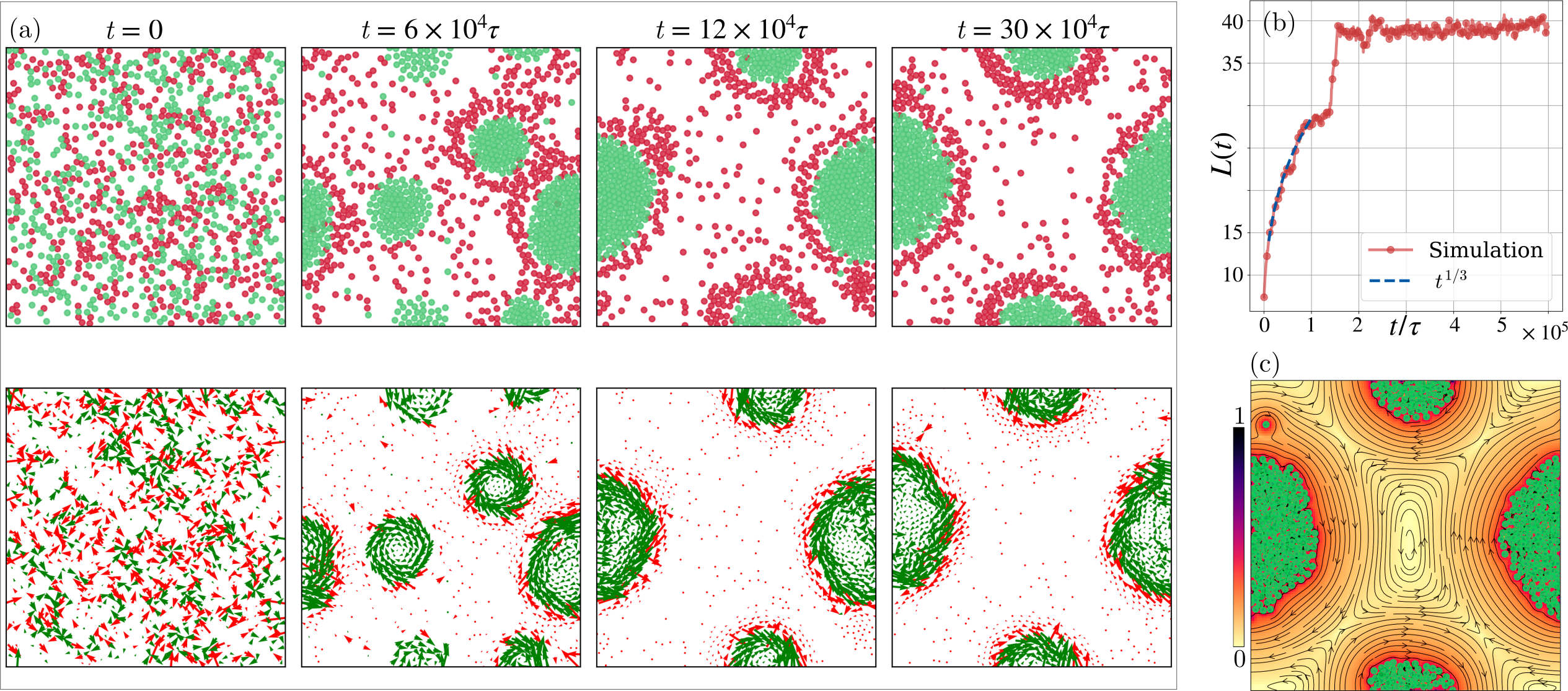}
    \caption{
    Snapshots (top row) and 
    vector plots of displacements (bottom row)  
    at distinct time points
     of active particles (green) and passive particles (red) is shown in panel (a). The vector plots shows a current of passive particles at the edge of cluster of active particle at later times. Note that this current is present even in absence of any boundary in the system. Here, we have considered a system of 1000 particles, $N_a=N_p=500$, $\phi=0.3$ and $\psi=0.8$. Movie III 
     in the SI confirms
     that these vortices appear for a larger system size of 5000 particles.
 (b): Length scale $L(t)$ of the coarsening domain - obtained using Eq.\eqref{eq:avgK} - as a function of time showing arrest in coarsening. 
      (c): streamlines of flow in the arrested state. For simplicity, only active particles have been shown as green circles. 
 The flow streamlines have been overlaid on the pseudo-color plot of the normalized logarithm of the flow speed.
    }
    \label{fig:ringArrest}
\end{figure*}

While the length scale gives us an idea of the underlying dynamics, the nature of fluid mediated coupling is given by the velocity distribution of passives. The velocity probability density functions (PDF) of passive particles immersed in an active chiral bath serve as a sensitive probe for the non-equilibrium energy injection scales . In contrast to the Maxwell-Boltzmann distributions characteristic of passive systems at thermal equilibrium, the velocity PDFs of passive tracers in this system, seen from Fig.\ref{vel_dist}, exhibit marked non-Gaussianity, typically characterized by "fat tails" that decay as exponential or power laws rather than quadratically. These heavy tails represent high-velocity events driven by the intermittent, "kicking" nature of the active spinners as they pass within the hydrodynamic near-field of the passive colloids. 
 Our system has long range hydrodynamic interactions which play a significant role in the phase separation. The velocities of passive particles, are thus, correlated and not independent, giving rise to the long-tailed nature of the distribution. The distribution also signifies the emergence of collective edge currents where passive particles are "entrained" by the coherent flow fields generated by synchronized spinner ensembles. 
Similar distributions as the one in Fig.\ref{vel_dist} are observed for all values of $\phi$ and $\psi$ where a phase separation dynamics is observed, see Fig.\ref{fig:phaseSep_4a}.

\subsection{Collective dynamics at the interface between active and passive domains}\label{sec:dyn_passive}
In this section we study the edge currents and their role in stabilising the arrested clusters. In Fig.\ref{fig:ringArrest}, we present vector plots of displacement of active particles (green) and passive particles (red) between two consecutive steps.
The  plot of displacement vector shows an edge current at the interface between the two phases .
 As evident from the figure, the particles in the bulk of the cluster do not undergo much displacement but at the edges they generate a vortical flow. This explains the formation of the ``ring" like structures.
  This suggests that an effective attraction has been induced between the active particles, due to the presence of the passives. The mechanism of phase separation is further substantiated by the flow around two passive particles as shown in Fig.\ref{fig:schematic}.(b) where the flow between the spinners is close to zero and the lines indicate how the passives responding to the flow prefer moving along the periphery of the active structures instead of moving into the bulk. Fig.\ref{fig:schematic}.(c) shows the path followed by a passive placed between two spinners, slightly off center. It is evident from the trajectory that the passive particle spends most of the time circling the structure formed by the actives. Such chiral behavior around the interface is also evident in the final snapshots of Fig.\ref{fig:ringArrest}, where circulation of passive particles (shown in red) is found around cluster of active particles (shown in green). 
  
  Interestingly, we find that the final phase remains arrested for the duration of the simulation due to chiral flows on the interfaces. This is also supported by the length scale plot in Fig.\ref{fig:ringArrest}(b). There is a cross over from growth of length scale
  to a plateau. The flow plot for the arrested state, given in Fig.\ref{fig:ringArrest}(c) also shows that the flow speed is higher near the active clusters and close to zero at the center. 
  We find arrested states only in the phase diagram where vortices are formed. Thus, they are sensitive to $\phi$. 
  A study of these arrested states from penetration length computed from the displacement plots as a function of activity would be studied in future.
  
 %
 
\subsection{Role of number asymmetry}
\label{subsec:asymm}
We have focused on 50:50 suspensions so far such that the number of active and passive particles is same in the system. In this section, we study the role of asymmetry in the number of active and passive particles.

\begin{figure}
\centering
\includegraphics[width=0.475\textwidth]{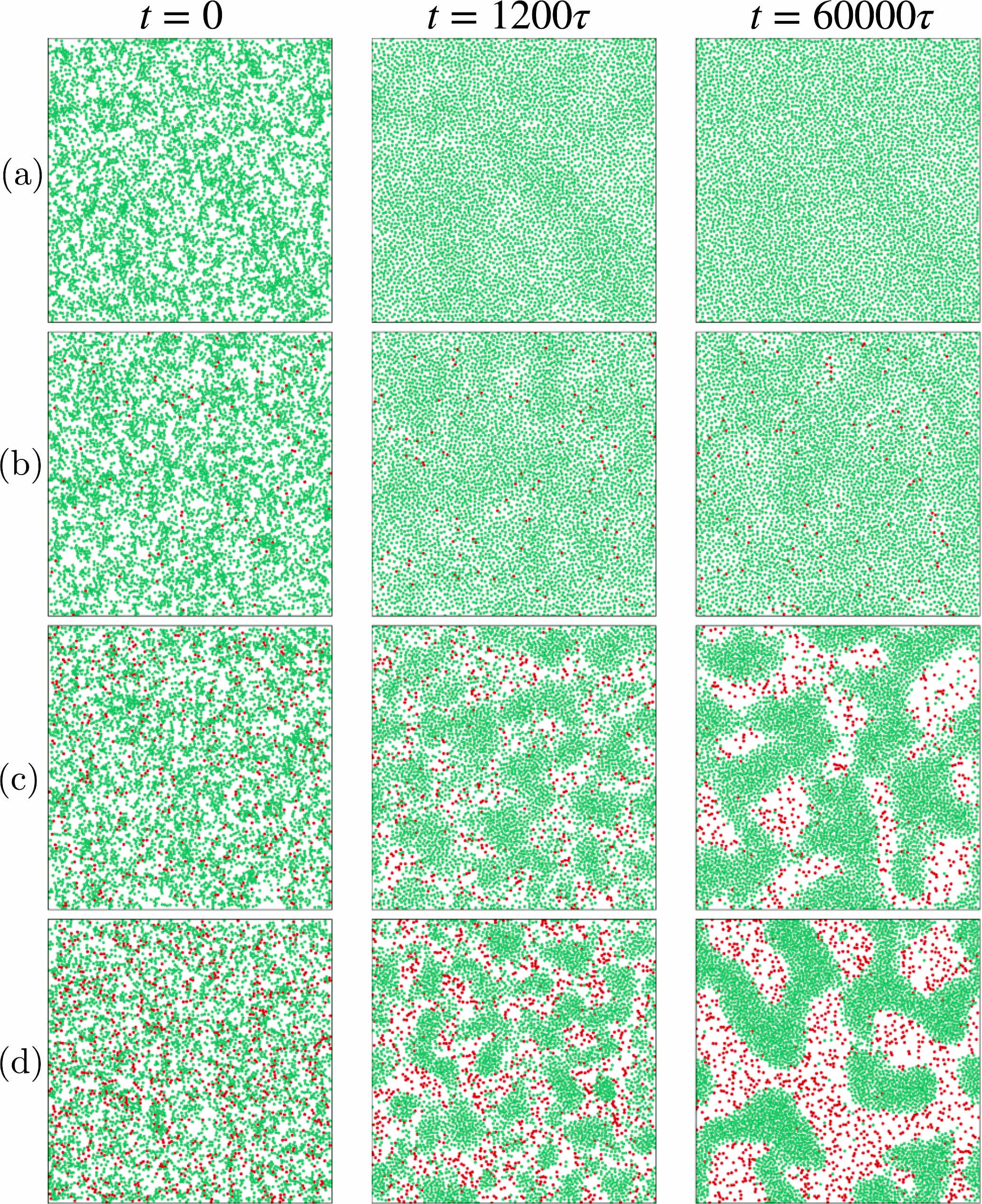}
\caption{Role of number asymmetry in temporal dynamics of a chiral system of $N=5000$ particles. Panel (a) shows a suspension of $N_a=5000$ active spinners only. Here $N_p=0$.  As seen from subsequent snapshots, there is no phase separation in this system. Panel (b) shows a system of $N_p=100$ passives and $N_a=4900$ actives. Panel (c) shows a suspension of $N_p=500$ passives and $N_a=4500$ actives. Panel (d) shows a suspension of 1000 passives and 4000 actives. All the panels shown are at $\phi=0.3$ and $\psi=1.2$}
\label{fig:numberasymm}
\end{figure}

Our results of this section are summarized in Fig.\ref{fig:numberasymm}.
In Fig.\ref{fig:numberasymm}a, we show that
a suspension of active spinners only, as observed from the snapshots this suspension does not show any clustering. We then eventually add passives to the suspension in panels (b), (c) and (d). we observe that with increase in number of passives in the suspension, phase separation sets in faster, showing that the addition of passives to the suspension does indeed introduce an attraction between the active spinners. We also note that the threshold value of $\phi$ increases for suspensions at lower passive to active ratio. 
It was observed to increase to around $\phi=0.3$ from 0.18 observed in case of 50:50 suspensions.
In appendix, we show the length scale plot of Fig.\ref{fig:numberasymmLvsT} for a suspension of only spinners shows there is an initial dip in the length scale plot. This is because the flow around the spinners 
and steric repulsion between the particles is making the 
distribution of the particles more uniform than the random initial condition used which had some small clusters of particles. Over time, the length scale saturates showing that no clustering is takes place. 

\section{Summary}\label{sec:summary}
In this study, we explore the dynamic behavior of a mixture comprising actively spinning particles, 
modeled as active spinners with fixed rotation axes, and passive tracer particles, all confined to a two-dimensional plane.
We show that a pure suspension of active spinners maintains long-term homogeneity, displaying no emergent structures. This stability arises from the symmetric cancellation of the self-generated hydrodynamic flows produced by the rotors' collective spinning motion. However, the introduction of passive particles dramatically alters this landscape by inducing a flow-mediated attraction between the spinners. 
 These interactions drive demixing and the system spontaneously organizes into a variety of striking morphologies determined primarily by the intensity of the chirality of the active particles and the local density of the mixture. 
 For 50:50 suspensions, the phase separation is clearly observed starting from total area fractions $\phi\approx 0.2$ and above, across a range of spinner activity levels ($\psi$). For mixtures at lower ratios of $N_p$ to $N_a$, no phase separation is observed at $\phi=0.2$ within the duration of simulation.


Particularly striking results emerge in 50:50 suspensions of spinners and passives, where we systematically vary the total area fraction $\phi$ and the spinner fraction $\psi$. These parameter sweeps reveal the evolution of unique, non-equilibrium morphological structures, including persistent vortices of passives around active clusters and macroscopic bands that span the system size. Such patterns highlight the potential for active matter systems to generate complex, topologically nontrivial phases through passive-mediated interactions, with implications for designing synthetic microswimmers and understanding biological active-passive mixtures like bacterial suspensions with inert colloids.

In this study, we have fixed the axis of rotation of the chiral active swimmer in a direction perpendicular to the monolayer distribution of particles. Here, we assume it to be strongly bottom-heavy as the orientation 
axis may change due to interactions with other particles.
Studying the collective behavior of swimmers when the orientation of the swimmers is allowed to change presents an exciting avenue for future work. A similar study of a suspension of polar particles with passive particles is also an interesting direction for future work. Study of binary suspensions under confining potentials to understand the interplay of activity and external potential is also a potential direction for future work.
In addition, we would like to note 
that
our current simulation is indeed based on the Ewald summation of the three-dimensional Oseen tensor (the Green's function for 3D Stokes flow) \cite{pozrikidis1992}. On the other hand, the particles are confined to a two-dimensional monolayer. 
We envisage that the underlying phenomenology will remain qualitatively similar in a strictly two-dimensional simulation, where the fluid flow is also two-dimensional. Because the active flow in our current setup already exhibits no out-of-plane components, the fundamental character and symmetries of the flow field are preserved. The transition to a truly two-dimensional system would primarily introduce quantitative changes, as the fluid velocity will attenuate differently as a function of distance.

\section*{Acknowledgments}
We thank two anonymous reviewers for comments, which has led to improvement of our manuscript.

\appendix

\section{Explicit form of the rigid body motion }\label{seappc:explicitRBM}
In this section, we provide explicit forms of RBM (rigid body motion) that can be used 
for dynamical simulations reported in this paper. 
In Eq.\eqref{eq:U1s4a} of the main text, we give the expression for kernels of the rigid body motion of particles. Here, we provide the explicit forms that have been used to run dynamical simulations. 
We first give the expression of the
term $U^{1s}_{\alpha\beta} $ in the index notation:
\begin{align*}
U^{1s}_{\alpha\beta,i} &=
\left(  1 + \frac{b^2}{3} \nabla^{2} \right)\,G_{ij}F^e_{\beta,j}.
\end{align*}
In the above, $G_{ij}$ is the Oseen tensor, which is the Green's function of Stokes equations with only boundary condition that the flow vanishes at infinity. 
The Oseen tensor $G_{ij}$ is given as \cite{pozrikidis1992}:
\begin{align}
8\pi \eta\,{G}_{ij}(\bm{r})= \frac{\delta_{ij}}{r} + \frac{r_i r_j}{r^{3}}   
=\left(\delta_{ij}\nabla^{2}-{\nabla_i}{\nabla_j}\right)r.
\label{eq:Oseen}
\end{align}
In periodic suspensions, one needs to perform lattice sum of the Oseen tensor from the periodic
images. To this end, we
compute the Ewald summation of the Oseen tensor, following 
Ref. \cite{beenakker1986ewald}. We first write:
\begin{alignat}{1}
\bm{G}(\bm{r}) & =\frac{1}{8\pi\eta}
\left(\bm{I}\,\nabla^{2}-\bm{\nabla}\bm{\nabla}\right)\big[
r\,\mathrm{\mathrm{erfc} }(\xi r)
+r\,\mathrm{erf }(\xi r)\big]
\label{eq:Gerf}
\end{alignat}
In the above, $\xi$ is a positive constant with dimensions of inverse length and we have defined
\begin{align}
\mathrm{erf}(q) = \frac{2}{\sqrt{\pi}} \int_0^q e^{-t^2}\mathrm{d}t,\quad 
\mathrm{erf}(q) + \mathrm{erfc}(q)=1.
\end{align}

Consider Eq.\eqref{eq:Gerf}, where 
the Green's function of Stokes flow 
is written in terms of parts decaying rapidly in real and Fourier spaces:
\begin{subequations}
    \label{eq:GerfExp}
\begin{align}
  8\pi\eta\,  \mathbf{G}^{\mathbb R}
(\mathbf{r}) &=  \left(\mathbf{I}\,\nabla^{2}-\boldsymbol{\nabla}\boldsymbol{\nabla}\right) r\, \mathrm{\mathrm{erfc} }(\xi r) 
,\\
8\pi\eta\, \mathbf{G}^{\mathbb F} (\mathbf{r}) &=\left(\mathbf{I}\,\nabla^{2} -\boldsymbol{\nabla}\boldsymbol{\nabla}\right) r\, \mathrm{\mathrm{erf} }(\xi r). 
\end{align}
\end{subequations}
In the above, 
the function $\mathbf{G}^{\mathbb R}(\mathbf{r})$ decays rapidly in real space, while the function $\mathbf{G}^{\mathbb F}(\mathbf{r})$ decays slowly in the real space (and thus rapidly in the Fourier space).
Using this information, the summation of the Green's function
can be accelerated using Ewald summation in both real and Fourier space \cite{beenakker1986ewald}.
To this end, we define the Fourier transform of a function $f(\mathbf{r})$
as:  
\begin{align}
    {\hat f}(\mathbf{k})&=\int\,
    f(\mathbf{r})e^{-i\mathbf{k}\cdot\mathbf{r}}d\mathbf{r},\qquad
f(\mathbf{r})=\int \hat {f}(\mathbf{k})e^{i\mathbf{k}\cdot\mathbf{r}} 
\frac{d\mathbf{k}}{(2\pi)^{3}}.\nonumber
\end{align}
Here, $\mathbf{k}$ is the wave-vector. 
Using the above, 
the Fourier transform of the Oseen tensor is:
\begin{align}
    \hat{G}_{ij}^{\mathrm{ }}(\mathbf{k})= 
    \frac{1}{\eta k^2}
    \left(
    \delta_{ij} -  \hat{k}_{i}\hat{k}_{j}
    \right).    
\end{align}
Here $\hat k_i = k_i/k$, while $k=|\mathbf{k}|$. 
Using this expression in Eq.(\ref{eq:GerfExp}), the expressions for $\mathbf{G}^{\mathbb R}$
and $\hat{\mathbf{G}}^{\mathbb F}$ becomes:
\begin{subequations}
\begin{align}
\label{eq:GrGf}
8\pi\eta \,\mathbf{G}^{\mathbb R}(\mathbf{r})
&=A(\xi r)\frac{\mathbf{I}}{r}+B(\xi r)\frac{\hat{\mathbf{r}}\hat{\mathbf{r}}}{r},\\
8\pi\eta \,\hat{\mathbf{G}}^{\mathbb F}(\mathbf{k}) 
&=C( k/\xi)\,\cos(\mathbf{k}\cdot\mathbf{r})\,\left(\frac{\mathbf{I}-\hat{\mathbf{k}}\hat{\mathbf{k}}}{k^{2}}\right).
\end{align}
\end{subequations}
The constants $A(q)$ and $B(q)$ are given as
\begin{align*}
  A(q)&=
  \mathrm{erfc}(q)+\Lambda(q)\,(2q^{3}-3q) 
\\
B(q)&=
\mathrm{erfc}(q)+\Lambda(q)\,(q-2q^3) 
\end{align*}
Here, we defined: $\Lambda(q)=\frac{2e^{-q^2}}{\sqrt{\pi}}$.
The function $C(q)$ is given by
\begin{align}
C(q)&=\frac{1}{\eta \mathcal{V}}\left[1+\frac{1}{4}q^2+\frac{1}{8}q^4\right]e^{-\frac{1}{4}q^4}.\nonumber
\end{align}
Here, $\mathcal{V}$ is the volume of the simulation box.\\

We now use the above form of Oseen tensor to obtain explicit forms of RBM. We first obtain explicit expression for the $1s$ mode, which has been formally defined in Eq.\eqref{eq:U1s4a}. 
The explicit form is given as:
\begin{align*}
       U^{1s}_{\alpha\beta,i}
       &=    \frac{F^e_{\beta,i}}{ 8\pi\eta}
     \left(A(\xi {r}) \frac{\delta_{ij}}{{d}} + B(\xi {r})\frac{{r}_{i} {r}_{j}}{r^3}\right)  
        \\ &
     +    \frac{F^e_{\beta,i}}{ 8\pi\eta}
    \frac{b^2}{3} 
          \left(\tilde A(\xi {r}) \frac{\delta_{ij}}{r^3} + \tilde B(\xi r)\frac{r_{i} r_j}{r^5}\right)  
          \\ &
          +       
  F^e_{\beta,i} \left(  1 - \frac{b^2k^2}{3}  \right) \hat G^{\mathbb F}_{ij}(\bm{k})
  \end{align*}
The terms
$ \tilde A(q)$ and $ \tilde B(q)$ are given as    
\begin{align}
    \tilde A(q) &=  2\mathrm{\mathrm{erfc}}(q)+
\Lambda (q)
\left[2q + 28 q^3  -40 q^5 + 8 q^7\right] \nonumber
\\
\tilde B (q)&=  -6\mathrm{\mathrm{erfc}}(q) + \Lambda(q) 
\left[ -6q - 4q^3 + 32 q^5-8 q^7 \right]     \nonumber
\end{align}


We now give the explicit form of RBM due to the $4a$ mode of the active slip, which has been formally defined in Eq.\eqref{eq:U1s4a}. 
In index notation, it is given as \cite{deb2025ewald}:
\begin{align}
      U^{4a}_{\alpha\beta,i}&=    \tilde\nabla_{p}\tilde\nabla_{u}\left(\epsilon_{lmj}\tilde\nabla_m G_{ij} \right) \Gamma_{\beta, lpu}
\end{align}
Explicit form is:
\begin{align*}
  U^{4a}_{\alpha\beta,i}&=  \frac{1}{8\pi\eta} \bigg[
   - \frac{\Upsilon(\xi r)}{r^7}\Big( \epsilon_{ijk} r_j  r_m r_n  \Big)\,
   \\
 &+  
    C(\tfrac{k}{\xi})\,\sin(\bm{k}\cdot \bm{r})
    \epsilon_{i jk} \frac{k_{j}k_{m} k_{n}} {k^2}
 \bigg]\,\Gamma_{\beta,jmp} 
\end{align*}
where $\Upsilon$ is given by
\begin{align}
\Upsilon &= 30\mathrm{\mathrm{erfc}}(q)
 + \Lambda(q)\Big[ 6q  +32 q^3 - 32 q^5 - 80q^7 + 16q^9\Big] \nonumber 
\end{align} 
Here, we have used the definition: $\Lambda(q)=\frac{2e^{-q^2}}{\sqrt{\pi}}$.

\begin{figure}[!t]
\centering
\includegraphics[width=0.45\textwidth]{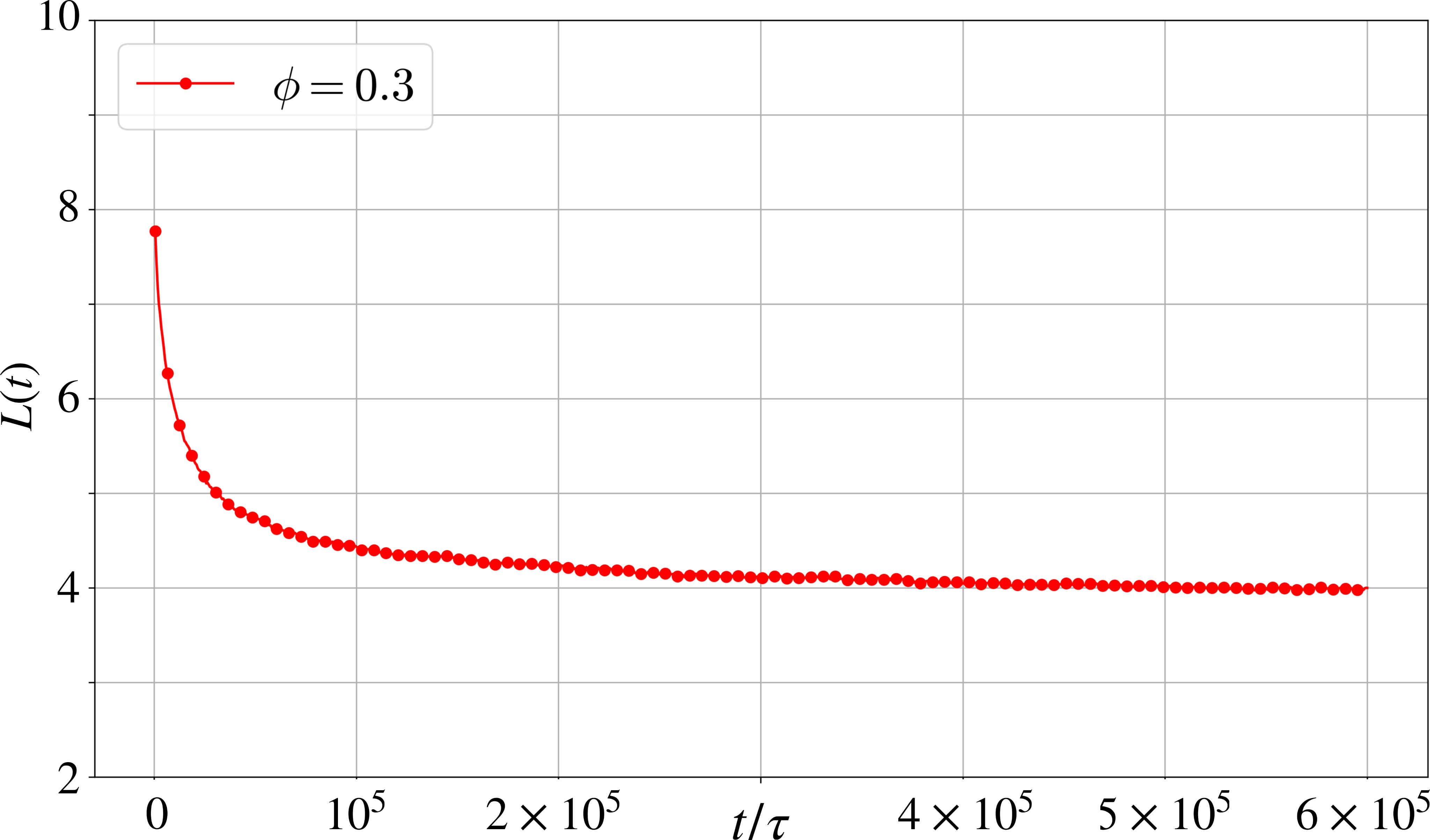}
\caption{
Time evolution of length scale $L(t)$ for a system of 2000 active particles. Here $N_p=0$, $\phi=0.3$ and $\psi=1.2$. Circles are from simulation, while the curve is a guide to the eye. 
It can be seen that the system reaches a homogeneous state. This is similar to Fig(\ref{fig:numberasymm})a but done for a smaller system of $N=2000$ so that simulation runs
for long time to show that system remains uniform in a system of only active spinners.}
\label{fig:numberasymmLvsT}
\end{figure}

\section{Simulation details}
\label{app:sim_details}
In this section, we provide simulation details and parameters used for dynamical simulations reported in this paper. 
The above forms of rigid body motion 
given in section \ref{seappc:explicitRBM} was 
simulated using the 
PyStokes library \cite{singh2020PyStokes}.
We note that the functionality for RBM computation in periodic domain was already existing in PyStokes and we adapted it to study the simulation of active-passive mixtures in this paper. 
We consider a simulation box of 
$L\times L\times L$. We initialise the system (a $50:50$ mixture of actively spinning and passive particles) in a way that the active particles and passive particles are in the $xy$-plane at $z=L/2$. 
All the active and passive particles are confined to a two-dimensional plane . 
The radius $b$ of each particle is 1.
The initial distribution of the particle is random and fully mixed. Thus, the whole system of active and passive particles form a monolayer at $z=L/2$. These particles remain in the monolayer throughout the simulation because the flow generated by the self-rotation of active particles does not lead to any motion out-of-plane.
Although the particles are moving in a two-dimensional space, their  motion is affected by the full three-dimensional hydrodynamic interactions. 
 We run the simulation for a total time of $1000s$ in all the cases. In figure \ref{fig:ringArrest} (a) the simulation was run for $10000s$. For the overlap precluding body force, the value $k^e$ is 60.  
 In our simulation, we set the optimal value of $\xi = \sqrt{\pi}/L$, 
such that the Ewald sum converges rapidly \cite{nijboer1957calculation,beenakker1986ewald}.

\section{Details of Supplementary Movies}
\label{supplemntary}

 Movies corresponding to four sets of parameters seen in Fig.\ref{fig:phaseSep_4a} along with movies for Fig. \ref{fig:ringArrest}(a) 
 have been provided as supplementary movies. The details are as follows:
 \begin{itemize}
     \item Movie 1 corresponds to Fig.\ref{fig:phaseSep_4a}. It is done for parameters values of $\psi=1.2$ and four values of $\phi$, which are: (0.1,0.2,0.4,0.7).
     \item Movie 2 corresponds to Fig.\ref{fig:ringArrest}.
     \item Movie 3 shows a system of $N_a=2500$ and $N_p=2500$ at $\phi=0.3$ and $\psi=0.8$. It confirms that the vortices are observed at larger system sizes too.
     \item Movie 4 shows a system of $N_a=2500$ and $N_p=2500$ at $\phi=0.7$ and $\psi=1.2$. It confirms that the bands are observed at larger system sizes too.
\end{itemize}


%

\end{document}